\documentclass[12pt]{article}
\usepackage{amsmath}
\usepackage{amsfonts}
\usepackage{amssymb}
\usepackage{srcltx}

\newtheorem{theorem}{Theorem}
\newtheorem{proposition}[theorem]{Proposition}

\newtheorem{corollary}[theorem]{Corollary}
\newtheorem{definition}[theorem]{Definition}

\newtheorem{remark}[theorem]{Remark}
\newtheorem{example}[theorem]{Example}

\begin{document}
\noindent
{\Large {\bf Polynomial tau-functions  of  the $n$-th Sawada-Kotera hierarchy}}
\vskip 9 mm
\begin{minipage}[t]{70mm}
{\bf Victor Kac}\\
\\
Department of Mathematics,\\
Massachusetts Institute of Technology,\\
Cambridge, Massachusetts 02139, U.S.A\\
e-mail: kac@math.mit.edu\\
\end{minipage}\qquad
\begin{minipage}[t]{70mm} 
{\bf Johan van de Leur}\\
\\
Mathematical Institute,\\
Utrecht University,\\
P.O. Box 80010, 3508 TA Utrecht,\\
The Netherlands\\
e-mail: J.W.vandeLeur@uu.nl
\end{minipage}
\\
\
\\
\begin{abstract}
We give a review of the B-type Kadomtsev-Petviashvili (BKP)  hierarchy,
and  find all polynomial tau-functions of the $n$-th reduced BKP hierarchy (=$n$-th  Sawada-Kotera hierarchy).
The name comes from the fact that for $n=3$ the simplest equation of the hierarchy  is the famous Sawada-Kotera equation.
\end{abstract}
\section{Introduction}
The three most famous hierarchies of Lax equations on one function $u$ are the Korteweg-de Vries (KdV) hierarchy,  the Kaup-Kupershmidt hierarchy,  and the Sawada-Kotera hierarchy. The Lax operators are, respectively,
\begin{align}\label{1}
{\cal L}=\partial^2 +u,\\
{\cal L}=\partial^3 +u\partial +\frac12 u',\label{2}\\
{\cal L}=\partial^3 +u\partial.\label{3}
\end{align}

Let $t=(t_1,t_2,t_3,\ldots)$
and
$
\tilde t= (t_1,t_3,t_5,\ldots)$.
Recall that the Kadomtsev-Pertviashvili (KP) hierarchy   is the following hierarchy of Lax equations on the pseudodifferential operator
$L(t,\partial)=\partial +u_1(t)\partial^{-1}+ u_2(t)\partial^{-2}+ \cdots $ 
 in $\partial=\frac{\partial }{\partial t_1}$ \cite{10a}:
\begin{equation}
\label{4}
\frac{\partial L(t,\partial)}{\partial t_k}=\left[
\left(
L(t,\partial)^k
\right)_{\ge 0}, L(t,\partial)
\right],\quad k=1,2,\ldots.
\end{equation}
The KdV hierarchy is the second reduced KP hierarchy, meaning that one imposes the following constraint on $L(t,\partial)$:
\begin{equation}
\label{5}
{\cal L}(t,\partial)=L(t,\partial)^2\quad\mbox{is a differential operator.}
\end{equation}
Then this operator is given by \eqref{1},     with $u(t)=2u_1(t)$, and the KP hierarchy \eqref{4} reduces to the KdV hierarchy
\begin{equation}
\label{6}
\frac{\partial{\cal  L}(t,\partial)}{\partial t_k}=\left[
\left(
{\cal L}(t,\partial)^{\frac{k}2}
\right)_{\ge 0}, {\cal L}(t,\partial)
\right],\quad k=1,3, 5,\ldots.
\end{equation}
For even $k$ this equation is trivial, ${\cal L}(t,\partial)$ is given by \eqref{1},  and equation \eqref{6} for $k=3$ is the KdV equation \cite {DJKM2}.

Recall that, in order to construct solutions of the KP hierarchy and the reduced KP hierarchies one introduces the tau-function $\tau(t)$, defined by \cite{10a}, \cite{DJKM2}:
\begin{equation}
\label{7}
L(t,\partial)=P(t,\partial)\circ \partial\circ P(t,\partial)^{-1},
\end{equation}
where $P(t,\partial)$ is a pseudodifferential operator,  with the symbol
\begin{equation}
\label{8}
P(t,z)=\frac1{\tau(t)}\exp \left(-\sum_{k=1}^\infty \frac{z^{-k}}{k}\frac{\partial}{\partial t_k} \right)(\tau(t)).
\end{equation}
The tau-function has a geometric meaning as a point on an infinite-dimensional Grassmannian,  and in \cite{10a} Sato showed that all Schur polynomials are tau-functions of the KP hierarchy.  Recently,  all polynomial  tau-functions of the KP hierarchy and its $n$-reductions have been constructed in   \cite{7a} (see also \cite{KRvdL}).

The CKP hierarchy (KP hierarchy of type C)  can be constructed by making use of the KP hierarchy, and assuming the additional constraint 
$L(\tilde t,\partial)^*=-L(\tilde t,\partial)$
(see e.g. \cite{8a} for details).
Its 3-reduction is defined by the constraint ${\cal L}(\tilde t, \partial)=L(\tilde t, \partial)^3$ is a differential operator,
and the corresponding hierarchy is
\begin{equation}
\label{9}
\frac{\partial{\cal  L}(\tilde t,\partial)}{\partial t_k}=\left[
\left(
{\cal L}(\tilde t,\partial)^{\frac{k}3}
\right)_{\ge 0}, {\cal L}(\tilde t,\partial)
\right],\quad k=1,3, 5,\ldots, k\not\in 3\mathbb Z,
\end{equation}
where $\cal L$ is given by \eqref{2}.  The simplest non-trivial of these equations  arizes  for $k=5$, and it is the Kaup-Kupershmidt equation.
All polynomial tau-functions of \eqref{9} (and all $n$ reductions of the CKP hierarchy) have been constructed in \cite{8a}.

In the present paper we construct all polynomial tau-functions of the $n$-reduced BKP hierarchies (KP hierarchy of type B). These are hierarchies of Lax equations on the differential operator
\begin{equation}
\label{10}
{\cal L}(\tilde t, \partial)=\partial^n+ u_{n-2}(\tilde t)\partial^{n-2}+\cdots+u_{1}(\tilde t)\partial,
\end{equation}
satisfying the constraint
\begin{equation}
\label{11}
{\cal L}(\tilde t,\partial)^*=(-1)^n\partial^{-1} {\cal L}(\tilde t,\partial) \partial.
\end{equation}
The $n$-th reduced BKP hierarchy is 
\begin{equation}
\label{12}
\frac{\partial{\cal  L}(\tilde t,\partial)}{\partial t_k}=\left[
\left(
{\cal L}(\tilde t,\partial)^{\frac{k}n}
\right)_{\ge 0}, {\cal L}(\tilde t,\partial)
\right],\quad k=1,3, 5,\ldots, k\not\in  n\mathbb Z.
\end{equation}
We call it the $n$-th Sawada-Kotera hierarchy,   since for $n=3$,  $\cal L$ is given by \eqref{3}, and for $k=5$, equation \eqref{12} is the Sawada-Kotera equation  \cite{SK} (see equation \eqref{Sb7}).

In the present paper, using the description of polynomial tau-functions of the BKP hierarchy \cite{KvdLB2}, \cite{KRvdL} (see Theorem \ref{Theo}), we find all polynomial tau-functions for the $n$-th Sawada-Kotera hierarchies (see Theorem 
\ref{Theo reduced}), and,    in particular, for the Sawada-Kotera hierarchy (see Corollary \ref{cor}).

\section{The   BKP hierarchy and its polynomial tau-functions}
\label{Sec2}
In this section we recall the construction of the BKP hierarchy \cite{DJKM3} and description of its polynomial tau-functions from  \cite{KvdLB2}
and  \cite{KRvdL}.

Following Date, Jimbo, Kashiwara and Miwa \cite{DJKM3} (see also \cite{KvdLB2} for details) we introduce the BKP hierarchy in terms of the so called twisted neutral fermions $\phi_i$, $i\in \mathbb Z$, which are generators of a Clifford algebra over $\mathbb C$, satisfying the following anti-commutation relatons
\begin{equation}
\phi_i\phi_j+\phi_j\phi_i=(-1)^i \delta_{i,-j}.
\end{equation} 
Define a  right (resp.  left) irreducible module $F=F_r$ (resp. $F_l$) over this algebra  by the following action  on the vacuum vector $|0\rangle$  (resp. $\langle 0|$)  
\begin{equation}
\phi_0|0\rangle=\frac1{\sqrt 2}|0\rangle,\quad
\phi_j|0\rangle =0\qquad 
\left(\mbox{resp. }
\langle 0|\phi_0=\frac1{\sqrt 2}\langle 0|,\quad 
\langle 0|\phi_{-j}=0
\right),
\quad \mbox{for }j>0.
\end{equation}
The quadratic elements 
\[
\phi_j\phi_k-\phi_k\phi_j
\ \mbox{for } j, k\in\mathbb{Z}, \  j>k ,
\]
form a  basis  of   the infinite-dimensional Lie algebra $so_{\infty, odd}$ over $\mathbb C$.  Let $SO_{\infty, odd}$ be the corresponding Lie group.
We proved in \cite[Theorem 1.2a]{KvdLB}    that a non-zero element
  $\tau\in F$ lies in this Lie  group orbit of     the vacuum vector $|0\rangle$ if and only if  it satisfies  the BKP hierarchy in the fermionic picture,    i.e.,    the following equation in $F\otimes F$:
\begin{equation}
\label{S}
\sum_{j\in\mathbb{Z}} (-1)^j \phi_j\tau \otimes \phi_{-j}\tau=\frac12 \tau\otimes\tau.
\end{equation}

Non-zero elements of $F$, satisfying \eqref{S},
are called tau-functions of the BKP hierarchy
in the fermionic picture,.

The group $SO_{\infty, odd}$ consists of elements  $G$ leaving  the symmetric bilinear form 
\begin{equation}
\label{bil}
(\phi_j,\phi_k)=(-1)^j\delta_{j,-k}
\end{equation}
on $F$
invariant, i.e.
\begin{equation}
\label{bil2}
(G\phi_j,G\phi_k)=
(\phi_j,\phi_k).
\end{equation}
Stated differently,  

\begin{equation}
\label{5a}
G\phi_kG^{-1}= \sum_{j_\in \mathbb Z} a_{jk}\phi_k\ \mbox{(finite sum) with }\sum_{j \in \mathbb Z}(-1)^j
a_{jk} a_{-j\ell}=(-1)^k\delta_{k,-\ell}.
\end{equation}

The group orbit of the vacuum vector is the disjoint union of Schubert cells  (see Section 3 of \cite{KvdLB2} for details). 
These cells are  parametrized by the strict partitions  $\lambda=(\lambda_1,\lambda_2,
\ldots,\lambda_k)$, with $\lambda_1>\lambda_2>.\cdots>\lambda_k>0$.
Namely, the cell, attached to the partition $\lambda$ is 
\begin{equation}
\label{orbit}
C_\lambda=\{
v_1v_{2}
\cdots v_{k-1} v_k|0\rangle|\,   v_j=\sum_{i\ge -\lambda_j} a_{ij}\phi_i\  \mbox {(finite sum) with }a_{-\lambda_j,j}\ne 0 \}.
\end{equation}
An element $\tau\in C_\lambda $
corresponds  to the following  point in the maximal isotropic Grassmannian  (i.e.  a maximal isotropic   subspace of $V=\bigoplus_{j\in \mathbb Z}{\mathbb C}\phi_j$):
\begin{equation}
\label{ann}
{\rm Ann}\, \tau =\{ v\in V|\, vv_1v_{2}
\cdots v_{k-1} v_k|0\rangle=0\}.
\end{equation}
For instance
$
{\rm Ann}\, |0\rangle={\rm span}\{ \phi_1, \phi_2,\ldots\}.
$

Using  the bosonization of the equation \eqref{S}, one obtains a hierarchy of differential equations on $\tau$ (\cite {DJKM3}, \cite {JM},  \cite {KRvdL} or    \cite[Section 3]{KvdLB2}).  This bosonization is an isomorphism $\sigma$ between the spin module $F$ and the polynomial algebra $B=\mathbb C[\tilde t]=\mathbb C[t_1,t_3,t_5,\ldots]$.  Explicitly,  introduce the twisted neutral fermionic field 
\[
\phi(z)=\sum_{j\in\mathbb Z} \phi_j z^{-j},
\]
and the bosonic field
\[
\alpha(z)=\sum_{j\in\mathbb Z}\alpha_{2j+1}z^{-2j-1}= \frac12 :\phi(z)\phi(-z):,
\]
where the normal ordering 
$:\quad :$ is defined by
\[
:\phi_j\phi_k:=\phi_j\phi_k  -\langle  0|\phi_j\phi_k|0\rangle;
\]
equivalently: $:\phi_j\phi_k:=\phi_j\phi_k$ if $j\le k$ and $=-\phi_k\phi_j $ if $j>k$, except when $j=k=0$, then 
$:\phi_0\phi_0:=0$.
The operators $\alpha_j$ satisfy the commutation relations of the  Heisenberg Lie algebra 
\begin{equation}
\label{alpha-com}
[
\alpha_j,\alpha_k ]=\frac{j}2\delta_{j-k},
\quad \alpha_i|0\rangle=\langle 0|\alpha_{-i}=0,\quad \mbox{for } i>0,
\end{equation}
and its representation on $F$ is irreducible \cite[Theorem 3.2]{KvdLB}. 
Using this, we obtain
a vector space
isomorphism  $\sigma:F\to B$,  uniquely defined
by the following relations:
\begin{equation}
\label{sigmaalpha}
\sigma(|0\rangle)=1,\quad
\sigma\alpha_j\sigma^{-1}=\frac{\partial}{\partial t_j},\quad 
\sigma\alpha_{-j}\sigma^{-1}=\frac{j}2 t_j,\quad \mbox{for }j>0\ \mbox{odd}.
\end{equation}
Explicitly \cite[Section 3.2]{KvdLB} :
\begin{equation}
\label{sigmaphi}
\sigma\phi(z)\sigma^{-1}=
\frac1{\sqrt 2}
\exp \sum_{j=1}^\infty t_{2j-1}z^{2j-1}\exp \sum_{j=1}^\infty -2 \frac{\partial}{\partial t_{2j-1}}
\frac{z^{-2j+1}}{2j-1}
.
\end{equation}
Since  \eqref{S}  can be rewritten as 
\[
{\rm Res}_z\, \phi(z)\tau\otimes \phi(-z) \tau \frac{dz} z=\frac12 \tau\otimes \tau,
\]
under the isomorphism $\sigma$   the
equation \eqref{S}   turns into:
\begin{equation}
\label{Sb}
{\rm Res}_z 
e^{ \sum_{j=1}^\infty (t_{2j-1}-t'_{2j-1})z^{2j-1}}
e^{ \sum_{j=1}^\infty 2
\left(  \frac{\partial}{\partial t'_{2j-1}}-\frac{\partial}{\partial t_{2j-1}}\right)
\frac{z^{-2j+1}}{2j-1}}\tau(\tilde t)\tau(\tilde t')
\frac{dz}z
=\tau(\tilde t)\tau(\tilde t'),
\end{equation}
where $\tilde t=(t_1,t_3,t_5,\ldots)$ and   $\tilde t'=(t_1',t_3',t_5',\ldots)$.   Therefore, $ \tau(\tilde t)$ is the vacuum expectation value 
\begin{equation}
\label{vacuum expectation}
\tau(\tilde t)=\sigma\tau\sigma^{-1}=\langle 0| e^{ \sum_{j=1}^\infty t_{2j-1}\alpha_{2j-1}}\tau.
\end{equation}
Furthermore,    by making a change of variables, as in \cite[page 972]{JM},  viz.
$t_{2k-1}=x_{2k-1}-y_{2k-1}$ and $t'_{2k-1}=x'_{2k-1}-y'_{2k-1}$,  and using the elementary Schur polynomials $s_j(r)$, which are defined by
\begin{equation}
\label{Schur}
\exp\sum_{k=1}^\infty r_kz^k=\sum_{j=0}^\infty s_j(r)z^j,
\end{equation}
we can rewrite \eqref{Sb}, where we assume $x_{2k}=y_{2k}=0$:
\begin{equation}
\label{Sb2}
\sum_{j=1}^\infty s_j(-2 \tilde y)s_j(2\tilde\partial_y)\tau(\tilde x-\tilde y)\tau(\tilde x+\tilde y)=0,
\end{equation}
where
$\tilde y=(y_1,0,y_3,0,\ldots)$ and 
$\tilde \partial_y=(\frac{\partial}{\partial y_1},0, \frac13\frac{\partial}{\partial y_3}, 0,\frac15\frac{\partial}{\partial y_5}, \ldots)$.
Using Taylor's formula we thus obtain the BKP hierarchy of Hirota bilinear equations  \cite[page 972]{JM},:
\begin{equation}
\label{Sb3}
\sum_{j=1}^\infty s_j(-2 \tilde y)s_j(2\tilde\partial_u)\exp\sum_{j=1}^\infty   y_{2j-1}\frac{\partial}{\partial  u_{2j-1}}
\tau(\tilde x-\tilde u)\tau(\tilde x+\tilde u)\Big|_{\tilde u=0}=0.
\end{equation}
Using the notation $p(D)f\cdot g=p(\frac{\partial}{\partial u_1},\frac{\partial}{\partial u_2},\ldots) f(\tilde x+\tilde u)g(\tilde x-\tilde u)\Big|_{\tilde u=0}$,  this turns into
\begin{equation}
\label{Sb4}
\sum_{j=1}^\infty s_j(-2\tilde y)s_j(2\tilde D)e^{\sum_{j=1}^\infty y_{2j-1}D_{2j-1}}
\tau\cdot \tau =0.
\end{equation}

The simplest equation in this hierarchy is
\cite[Appendix 3]{JM}:
\begin{equation}
\label{Sb5}
(D_1^6-5D_1^3D_3-5D_3^2+9D_1D_5)\tau\cdot\tau=0.
\end{equation}
If we assume that our tau-function does not depend on $t_3$, then this gives 
\begin{equation}
\label{Sb6}
(D_1^6 +9D_1D_5)\tau\cdot\tau=0.
\end{equation}
Letting  $x=t_1$,  $t=\frac19 t_5$, and  
\begin{equation}
\label{17a}
u(x,t)= 2\frac{\partial^2 \log \tau(x,t)}{\partial x^2},
\end{equation}
  and viewing  the remaining $t_j$ as parameters,
 equation \eqref{Sb6} turns into the famous Sawada-Kotera equation \cite{SK}:
\begin{equation}
\label{Sb7}
u_t+15(uu_{xxx}+u_x u_{xx}+3u^2u_x)+u_{xxxxx}=0.
\end{equation}

Another approach is by using the wave function,  see \cite[page 345]{DJKM3},
\begin{equation}
\label{wave}
\begin{aligned}
w(\tilde t,z)&=\frac1{\tau(t)}
\exp  \sum_{j=1}^\infty t_{2j-1}z^{2j-1}
\exp {- \sum_{j=1}^\infty 2
    \frac{\partial}{\partial t_{2j-1}} 
\frac{z^{-2j+1}}{2j-1}
}
\tau(t) \\
&=P(\tilde t,z)e^{ \sum_{j=1}^\infty (t_{2j-1})z^{2j-1}},
\end{aligned}
\end{equation}
where $P(\tilde t,z)=1+\sum_{j=1}^\infty p_j(\tilde t)z^{-j}$, and in particular 
\begin{equation}
\label{p1}
p_1(\tilde t)=-2\frac{\partial \log \tau(\tilde t)}{\partial t_1}.
\end{equation}
Letting $P(\tilde t,\partial)$  be the pseudodifferential operator in $\partial =\frac{\partial}{\partial t_1}$  with the symbol $P(\tilde t,z)$, 
equation \eqref{Sb} turns into 
\begin{equation}
\label{SbP}
{\rm Res}_z P(\tilde t,\partial)P(\tilde t', \partial')
e^{ \sum_{j=1}^\infty (t_{2j-1}-t'_{2j-1})z^{2j-1}}
\frac{dz}z
=1.
\end{equation}
Now,  using the fundamental  lemma, Lemma 1.1 of \cite{DJKM2} or   Lemma 4.1 of \cite{KLmult},  we  deduce  from \eqref{SbP}:
\begin{equation}
\label{Sato}
\begin{aligned}
&P(\tilde t,\partial)\partial^{-1} P(\tilde t,\partial)^*\partial =1,\\
&\frac{ \partial
P(\tilde t,\partial)}{\partial t_{2j-1}}
=- (P(\tilde t,\partial)\partial^{2j-1}P(\tilde t,\partial)^{-1}\partial^{-1})_{<0}\partial P(\tilde t,\partial),\quad j=1,2,\ldots.
\end{aligned}
\end{equation}

Next, introducing the Lax operator 
\[
L(\tilde t,\partial)=P(\tilde t,\partial)\partial P(\tilde t,\partial)^{-1}= \partial +u_1(\tilde t)\partial^{-1}+u_2(\tilde t)\partial^{-2}+\cdots
,
\]
we deduce from \eqref{Sato} that $L$ satisfies \cite{DJKM3}
\begin{equation}
\label{Lax}
\begin{aligned}
L(\tilde t,\partial)^*&=-\partial^{-1} L(\tilde t,\partial)\partial,\\
\frac{ \partial
L(\tilde t,\partial)}{\partial t_{2j-1}}&
=\left[\left(L(\tilde t,\partial))^{2j-1}\right)_{\ge 0},L(\tilde t,\partial)\right],\quad j=1,2,\ldots.
\end{aligned}
\end{equation}
Note that,  since $u_1(\tilde t)=-\frac{\partial p_1(\tilde t)}{\partial t_1}$ and  the fact that  $p_1(t)$ is given by \eqref{p1}, we find that 
\begin{equation}
u_1(\tilde t)=2\frac{\partial^2 \log \tau(\tilde t)}{\partial t_1^2},
\end{equation}
which explains the choice \eqref{17a} of $u(x,t)$  to obtain the Sawada-Kotera equation from   the Hirota bilinear equation \eqref{Sb6}.

To obtain the second equation of \eqref{Lax}, we use \eqref{Sato} and  the first equation of \eqref{Lax}, which is equivalent
(see \cite{DJKM3}, page 356)
 to the fact that 
$L(\tilde t,\partial)^{2j-1}$,  for $j=1,2,\ldots$, has zero  constant term.
Let us prove that the first equation of \eqref{Lax} indeed implies this fact. We have 
\[
L^k\partial^{-1}=(-\partial^{-1}L^*\partial)^k\partial^{-1}
=(-1)^k\partial^{-1}L^{*k} 
=(-1)^{k+1} (L^k\partial^{-1})^*.
\]
Now, using the fact that the constant term of $L^k$ is equal to 
\[
{\rm Res}_\partial L^k\partial^{-1}=-{\rm Res}_\partial(L^k\partial^{-1})^*={\rm Res}_\partial 
(-1)^{k+1} (L^k\partial^{-1})^*
=(-1)^k {\rm Res}_\partial L^k\partial^{-1},
\]
we find that the constant term of $L^k$ is zero whenever $k$ is odd.
\begin{remark}
Note that this also means that we can replace the second equation of \eqref{Lax} by,  cf.  \cite{Kup},
\[
\frac{ \partial
L(\tilde t,\partial)}{\partial t_{2j-1}}
=\left[\left(L(\tilde t,\partial))^{2j-1}\right)_{\ge 1},L(\tilde t,\partial)\right],\quad j=1,2,\ldots.
\]
In the formulation of Kupershmidt \cite{Kup},  this means that $L$ satisfies not only the KP equation for the odd times,  but also his formulation of the modified KP hierachy (only for the odd times).
\end{remark}

Next,  we describe polynomial tau-functions $\tau(t_1,t_3,\ldots)$ of the BKP hierarchy obtained in   \cite[Theorem 6]{KvdLB2} (see also \cite{KRvdL}). For that, given integers $a$ and $b$, $a>b\ge 0$,  let 
\begin{equation}
\label{chi}
\begin{aligned}
\chi_{a,b}(t,t')&=\frac 12 s_a(t')s_b(t)+\sum_{j=1}^b (-1)^j s_{a+j}(t')s_{b-j}(t),\\
\chi_{b,a}(t,t')&=- \chi_{a,b}(t,t'),  \quad  \chi_{a,a}(t,t')=0,
\end{aligned}
\end{equation}
and let $\chi_{a,b}(t,t')=0$ if $b<0$. Then

\begin{theorem}\label{Theo}  {\rm \cite[Theorem 6]{KvdLB2} }
All polynomial tau-functions of the BKP hierarchy \eqref{Sb}, up to a scalar
multiple,  are equal to  
\begin{equation}
\label{TBKP}
\tau_\lambda (\tilde t)=Pf\left(\chi_{\lambda_i,\lambda_j}(\tilde t+c_i,\tilde t+c_j)\right)_{1\le i,j\le 2n},
\end{equation}
where   $Pf$ is the Pfaffiann of a skew-symmetric matrix, $\lambda=(\lambda_1,\lambda_2,\ldots ,\lambda_{2n})$
 is an extended strict partition, i.e. $\lambda_1>\lambda_2>\cdots >\lambda_{2n}\ge 0$,  $\tilde t=(t_1,0,t_3,0,\ldots)$, $c_i=(c_{1i},c_{2i}, c_{3i}, \ldots)$, $c_{ij}\in\mathbb C$.
\end{theorem}
\begin{remark}
The connection between the set of strict partitions   and the extended strict partitions is as follows.  If $\lambda=(\lambda_1,\lambda_2,\ldots,\lambda_k)$ is a strict partition and  $k$ is even, then this partition is equal to the extended strict partition $\lambda$.  However, if $k$ is odd,  the Pfaffian of a $k\times k$ anti-symmetric matrix is equal to $0$,  hence in that case we extend $\lambda$ by the element $0$,  i.e.,  the corresponding extended strict partition is then 
$(\lambda_1,\lambda_2,\ldots,\lambda_k,0)$.
\end{remark}

\section{The $n$-th Sawada-Kotera  hierarchy and its polynomial tau-functions}
\label{Sec3}
As we have seen in Section \ref{Sec2},  a necessary condition for a tau-function to give a solution of the Sawada-Kotera equation is that $\frac{\partial \tau(\tilde t)}{\partial t_3}=0$.
This means that the tau-function lies in a smaller group orbit of the vacuum vector $|0\rangle$. Instead of the $SO_{\infty,odd}$ orbit of the vacuum vector $|0\rangle$, we consider the twisted loop group $G_3^{(2)}$, corresponding  to the affine Lie algebra $sl_3^{(2)}$, to obtain  the 3-reduced BKP hierarchy \cite{DJKM3}. 
More generally (see also \cite{DJKM3}), when $n=2k+1>1$ is odd,  the $2k+1$-reduced hierarchy is related to the twisted loop group  $G_{2k+1}^{(2)}$ corresponding to  the twisted affine Lie algebra
$sl_{2k+1}^{(2)}$.  When $n=2k>2$ is even one has the twisted loop group $G_{2k}^{(2)}$  corresponding  to the affine Lie algebra $so_{2k}^{(2)}$
 \cite{DJKM1}, 
\cite{KvdLB} (see \cite[Chapter 7]{Kacbook} for the construction of these Lie algebras).
Elements $G$ in this twisted loop group not only satisfy \eqref{5a}, which implies 
$\sum_{j \in \mathbb Z}(-1)^j
a_{kj} a_ {\ell,-j}=(-1)^k\delta_{k,-\ell}$,
 but also the $n$-periodicity condition $a_{i+n,j+n}=a_{ij}$.
This means that these group elements also commute with the operator 
$$\sum_{i\in\mathbb{Z} 
} 
(-1)^{pn-i}\phi_i \otimes \phi_{pn-i}, \quad \mbox{for }p=1,2,3,\ldots,$$
namely 
\[
\begin{aligned}
(G\otimes G) 
\sum_{i\in\mathbb{Z} 
} 
(-1)^{pn-i}
\phi_i \otimes \phi_{pn-i}
&=\sum_{i\in\mathbb{Z}}(-1)^{pn-i}
G\phi_i  G^{-1}\otimes G  \phi_{pn-i}G^{-1} G\\
&=\sum_{i,j,k\in \mathbb{Z}}
(-1)^{pn-i}
 a_{ji} a_{k,pn-i}\phi_j G\otimes \phi_k G\\
&=\sum_{i,j,k\in\mathbb{Z}} 
(-1)^{pn-i}
a_{ji} a_{k-pn,-i}\phi_j G\otimes \phi_k G\\
 &=
 \sum_{j\in\mathbb{Z}} (-1)^{pn-j}\phi_j
 G\otimes \phi_{pn-j}G\\
&=\sum_{j\in\mathbb{Z}
} 
(-1)^{pn-j}\phi_j\otimes \phi_{pn-j} (G\otimes G) 
.
\end{aligned}
\]
Since $\sum_{i\in\mathbb{Z} 
} 
(-1)^{pn-i}\phi_i|0\rangle \otimes \phi_{pn-i}|0\rangle=0$, we find that the elements $\tau$ in the  orbit of the vacuum vector of this twisted loop not only satisfy \eqref{S},  
but also satisfy the conditions
\begin{equation}
\label{Sn}
\sum_{j\in\mathbb{Z}} (-1)^{pn-j} \phi_j\tau \otimes \phi_{pn-j}\tau=0, \quad p=1,2,\ldots.
\end{equation}
This means  that $\tau(\tilde t)=\sigma(\tau)$ not only satisfies  \eqref{Sb}, 
but also   the  conditions
\begin{equation}
\label{Sbn}
{\rm Res}_z  z^{pn-1}
e^{ \sum_{j=1}^\infty (t_{2j-1}-t'_{2j-1})z^{2j-1}}
e^{ \sum_{j=1}^\infty 2
\left(  \frac{\partial}{\partial t'_{2j-1}}-\frac{\partial}{\partial t_{2j-1}}\right)
\frac{z^{-2j+1}}{2j-1}}\tau(\tilde t)\tau(\tilde t')
 dz
=0, \quad p=1,2,\ldots.
\end{equation}

From \eqref{Sbn} one deduces, using the fundamental Lemma \cite[Lemma 4.1]{KLmult} and the first equation of \eqref{Sato},  that 
\[
(P(\tilde t,\partial)\partial^{pn-1}P(\tilde t,\partial)^*)_{<0}=(P(\tilde t,\partial)\partial^{pn}P(\tilde t,\partial)^{-1}\partial^{-1})_{<0}=0. 
\]
Thus,
the Lax operator $L(\tilde t,\partial)$ satisfies 
\begin{equation}
\left(L(\tilde t,\partial)^{pn} \right)_{\le 0}=0,\quad p=1,2,\ldots .
\end{equation}
Hence ${\cal L}(\tilde t, \partial )={  L}(\tilde t, \partial )^n$ is a monic differential operator with zero constant term.
Moreover,  ${\cal L}(\tilde t, \partial )$ is equal to \eqref{10}, and, by the first formula of \eqref{Sato},
we have the relation    \eqref{11}.

Now,  if $n$ is odd,   one  can use the  the Sato-Wilson equation, i.e. the second equation of   \eqref{Sato},  to find that 
\[
\frac{\partial P(\tilde t,\partial)}{\partial t_{(2j-1)n}}=0,\quad j=1,2,\ldots.
\]
From this we find that the tau-function satisfies 
\begin{equation}
\frac{\partial \tau(\tilde t)}{\partial t_{(2j-1)n}}=\lambda_j \tau(\tilde t),\quad \lambda_j\in\mathbb C, \quad\mbox{for }j=1,2,\ldots.
\end{equation}
Since we consider only polynomial tau-functions, we find that for odd $n$:
\begin{equation}
\label{diftau}
\frac{\partial \tau(\tilde t)}{\partial t_{(2j-1)n}}=0,\quad  j=1,2,\ldots.
\end{equation}
If $n$ is even there is no such restriction, because the Sato-Wilson equation \eqref{Sato} is only defined for odd flows.
However,  the additional equations \eqref{Sbn} still hold and  give additional constraints on the tau-function.
\begin{proposition}
\label{prop4}
For $n$ odd, 
equation \eqref{diftau} for $j=1$ and the BKP hierarchy \eqref{Sb} on $\tau(\tilde t)$ are equivalent to \eqref{Sb} and \eqref{Sbn}.
\end{proposition}
{\bf Proof.}
We only have to show that \eqref{diftau} for $j=1$ and \eqref{Sb} imply  \eqref{Sbn}.  For this differntiate \eqref{Sb} by $t_n$  and use  \eqref{diftau}, this gives Equation  \eqref{Sbn} for $p=1$. Next differentiate \eqref{Sbn} for $p=1$ again by $t_n$ and use again \eqref{diftau};
this gives  \eqref{Sbn} for $p=2$, etc. \hfill$\square$
\\
\begin{remark}
If $n$ is odd. Proposition \ref{prop4} gives that a polynomial BKP tau-function is $n$-th Sawada-Kotera tau-function if and only if $\tau$ satisfies 
$\frac{\partial \tau}{\partial t_n}=0.$
\end{remark}
 Since $L$ satisfies the BKP hierarchy ${\cal L}=L^n$ also satisfies the BKP hierarchy.  
 For $n=3$,   assuming the constraint that $\cal L$ is a differential operator,   ${\cal L}$ is given by \eqref{3} and 
$\frac{\partial{\cal L}}{\partial t_3}=[({\cal L}^{\frac53})_{\ge 0}, {\cal L}]$ is the Sawada-Kotera equation \eqref{Sb7}.
This leads to the following definition.
\begin{definition}
Let 
$
{\cal  L}= \partial^n+u_{n-2}\partial^{n-2}+\cdots +u_1\partial
 $
 be a differential operator,  satisfying   \eqref{11}.
  The system of Lax equations 
      \begin{equation}
\label{Laxn}
\frac{ \partial
{\cal L}(\tilde t,\partial)}
{\partial t_{2j-1}}
=
\left[
\left(
{\cal L}(\tilde t,\partial)^{\frac{2j-1} n}
\right)_{\ge 0},{\cal L}(\tilde t,\partial)
\right],\quad j=1,2,\ldots,
\end{equation}
is called the $n$-th reduced BKP hierarchy or the $n$-th Sawada-Kotera hierarchy. For $n=3$ it is called the Sawada-Kotera hierarchy.
\end{definition}

The geometric meaning of equation \eqref{Sn} is that the space ${\rm Ann}\, \tau$ is invariant under the shift
$\Lambda_n$,
where
\begin{equation}
\label{Lambda}
\Lambda_n(\phi_i)=\phi_{i+n}.
\end{equation}

As in the $SO_{\infty,odd}$ case,  all polynomial tau-functions in this $n$-reduced case lie in some Schubert cell.
Such a Schubert cell has a ``lowest'' element $w_\lambda$,  for $\lambda$ a certain strict partition.  This element can be obtained from the vacuum vector  by the action of the Weyl group  corresponding to $G_n^{(2)}$.  
The  element 
\begin{equation}
\label{w lambda}
w_\lambda=\phi_{-\lambda_1}\phi_{-\lambda_2}\cdots \phi_{-\lambda_k}|0\rangle
\end{equation}
lies in the Weyl group  orbit of $|0\rangle$,    corresponding to $SO_{\infty,odd}$,
see   \cite{You}, however not all such elements lie  in the Weyl group orbit  of $|0\rangle$ for $G_n^{(2)}$.   
For this, consider
\begin{equation}
\label{ann w lambda}
{\rm Ann}\, w_\lambda={\rm span} \{\phi_{-\lambda_1},\phi_{-\lambda_2},\ldots ,\phi_{-\lambda_k}\}\oplus
{\rm span} \{\phi_i|\, i>0, i\ne \lambda_j, \ j=1,\ldots,k
\}.
\end{equation}
The element  $w_\lambda$ lies in the $G_n^{(2)}$ Weyl group  orbit    if and only if  ${\rm Ann}\, w_\lambda$ is 
invariant under the action of $\Lambda_n$,   which means that the ($\lambda_1+1$ shifted) set
\begin{equation}
\label{V lambda}
V_\lambda=\{  \lambda_1+\lambda_i+1|
, \  i=1, \ldots ,k\} \cup 
\{  \lambda_1- j+1 |\, 0<j<\lambda_1,\ j\ne \lambda_i\ \mbox{for } i=1,\ldots, k\} 
,
\end{equation}
must satisfy the $-n$ shift condition, i.e., 
\begin{equation}
\label{condition V lambda}
\mbox{if }\mu_j\in V_\lambda, \ \mbox{then }\mu_j-n \in V_\lambda\ \mbox{ or }\mu_j-n\le 0.
\end{equation}
Only the  elements $w_\lambda$,  for which the corresponding $V_\lambda$ satisfies condition 
\eqref{condition V lambda} lie in the $G_n^{(2)}$ group orbit.
\begin{example}
\label{Ex}
(a) For $n=2$ the only strict partition $\lambda$ that satisfies condition \eqref{condition V lambda} is $\lambda=\emptyset$.
\\
(b)For  $n=3$, the only strict partitions $\lambda$  that satisfy condition  \eqref{condition V lambda}
are 
\[
(3m+1, 3m-2, 3m-5,\ldots, 4,1)\ \mbox{and }(3m+2, 3m-1, 3m-4,\ldots, 5,2), \quad m\in \mathbb Z_{\ge0}.
\]
\end{example}
\begin{remark}
\label{remark a}
Note that  \eqref{condition V lambda} means that $\lambda$   is a strict partition which is the union of strict partitions   $(nm+a_i, n(m-1)+a_i,  ,\ldots, n+a_i,a_i)$, with $1\le a_i<n$ 
and  $1\le i<n$, such that  $a_j-n\ne -a_\ell$. In other words $a_j+a_\ell\ne n$.   Hence there are   at most $\left[\frac{n}2-1\right]$
such $a_i$.
\end{remark}

To  a strict partition  $\lambda$,  that satisfies condition \eqref{condition V lambda},  the corresponding Schubert cell is then obtained through the action on  a  $w_\lambda$  by an upper-triangular matrix in the group $G_n^{(2)}$.  This produces,  up to a constant  factor, elements
\begin{equation}
\label{v lambda}
v_\lambda=v_1v_2\cdots v_k|0\rangle,\ \mbox{where }v_j=\phi_{-\lambda_j}+\sum_{i\ge 1 -\lambda_j}  a_{ij}\phi_i \ \mbox{(finite sum)},
 \end{equation}
  and 
 \begin{equation}
 \label{v lambda2}
 (v_j,v_\ell)=0, \ \mbox{for } j,\ell=1,\ldots ,k,\ 
 \mbox{and if }\lambda_i=\lambda_j-n,\ \mbox{then } v_i=\Lambda_n(v_j).
 \end{equation}
 
 We first  express the constants $a_{ij}$ in terms of other constants  by letting
 \[
 a_{ij}=s_{i+\lambda_j}(c_{\overline{\lambda}_j}), \ \mbox{ where the $s_i$ are elementary Schur polynomials}.
\]
Here we use  that
\[
1+\sum_{i=1 -\lambda_j}^\infty a_{ij}z^{i+\lambda_j}=\exp\left(\sum_{k=1}^\infty c_{k,\overline{\lambda}_j}z^k\right),
\]
hence
for every  $\overline{\lambda}_j$, one can recursively obtain the $c_{k,\overline{\lambda}_j}$.   Since,  $a_{ij}=0$ for $i>>0$,  one only has  a finite number of  $c_{k,\overline{\lambda}_j}$. 
Thus
 \begin{equation}
\label{v lambda4}
 v_j=\phi_{-\lambda_j}+\sum_{i> -\lambda_j}  s_{i+\lambda_j}(c_{\overline{\lambda}_j})\phi_i,
 \end{equation}
where $c_{\overline{\lambda}_j}=(c_{1,\overline{\lambda}_j},c_{2,\overline{\lambda}_j},c_{3,\overline{\lambda}_j},\ldots
 )$.   Here $\overline{\lambda}_j={\lambda}_j\mod n$, which means that there are at most $\left[\frac n 2-1\right]$ of such infinite series of constants (see Remark \ref{remark a})  and  the $v_j$ satisfy the condition   \begin{equation}
 \label{v lambda3}
 \mbox{if }\lambda_i=\lambda_j-n,\ \mbox{then } v_i=\Lambda_n(v_j).
 \end{equation}
 We can now use the isomorphism $\sigma$ to calculate the bosonization of elements $v_\lambda$.
 For this we use  formula \eqref{vacuum expectation} and apply this to  $v_\lambda$ (which is given by \eqref{v lambda}
  with $v_j$ given by \eqref{v lambda4}).
Now using \eqref{sigmaalpha} and  \eqref{sigmaphi} and the fact that  
\[
  e^{\sum_{j=1}^\infty  t_{2j-1}  \frac{\partial}{\partial s_{2j-1}}}
e^{ \sum_{j=1}^\infty s_{2j-1}z^{2j-1}}=e^{ \sum_{j=1}^\infty (t_{2j-1}+s_{j-1})z^{2j-1}} e^{\sum_{j=1}^\infty  t_{2j-1}  \frac{\partial}{\partial s_{2j-1}}}
\]
we find that
\[
e^{ \sum_{j=1}^\infty t_{2j-1}\alpha_{2j-1}}\phi(z)e^{- \sum_{j=1}^\infty t_{2j-1}\alpha_{2j-1}}=
e^{ \sum_{j=1}^\infty t_{2j-1}z^{2j-1}}\phi(z).
\]
Thus, using \eqref{v lambda4}, 
  we find that   
\[
\begin{aligned}
v_j( \tilde t):=&e^{ \sum_{j=1}^\infty t_{2j-1}\alpha_{2j-1}}v_j e^{- \sum_{j=1}^\infty t_{2j-1}\alpha_{2j-1}}
\\
=&e^{ \sum_{j=1}^\infty t_{2j-1}\alpha_{2j-1}}(\phi_{-\lambda_j}+\sum_{i\ge 1 -\lambda_j} s_{i+\lambda_j}(c_{\overline{\lambda}_j})\phi_i)e^{- \sum_{j=1}^\infty t_{2j-1}\alpha_{2j-1}}\\
=&e^{ \sum_{j=1}^\infty t_{2j-1}\alpha_{2j-1}}
{\rm Res}\, \sum_{\ell=0}^\infty s_\ell (c_{\overline{\lambda}_j})z^{\ell-\lambda_j}\phi(z)
e^{- \sum_{j=1}^\infty t_{2j-1}\alpha_{2j-1}}\frac{ dz}z\\
=&{\rm Res}\, \sum_{\ell=0}^\infty s_\ell (c_{\overline{\lambda}_j})z^{\ell-\lambda_j}\phi(z)e^{ \sum_{j=1}^\infty t_{2j-1}z^{2j-1}}\frac{ dz}z\\
=&
{\rm Res}\,z^{-\lambda_j} e^{\sum_{i=1}^\infty c_{i,\overline{\lambda}_j}z^i+ \sum_{j=1}^\infty t_{2j-1}z^{2j-1}} \phi(z)\frac{ dz}z\\
=&
{\rm Res}\,z^{-\lambda_j}  
\sum_{k=0}^\infty  s_{k }(\tilde t+c_{\overline{\lambda}_j})z^k
 \sum_{i\in\mathbb Z}\phi_iz^{-i}\frac{ dz}z\\
=&\phi_{-\lambda_j}+\sum_{i\ge 1 -\lambda_j}  s_{i+\lambda_j}(\tilde t+c_{\overline{\lambda}_j})\phi_i
.
\end{aligned}
\]
Since $e^{ \sum_{j=1}^\infty t_{2j-1}\alpha_{2j-1}}|0\rangle =0$, we find that the corresponding tau-function is equal to the vacuum expectation value 
\begin{equation}
\label{tau vacuum exp}
\tau(\tilde t)=\langle 0|v_1(\tilde t)v_2(\tilde t)\cdots v_k(\tilde t)|0\rangle.
\end{equation}
If $k=2m$, then this is the  Pfaffian of a $2m\times 2m$ skew-symmetric matrix.    If $k=2m-1$, we use the fact hat 
\[
\langle 0|v_1(\tilde t)v_2(\tilde t)\cdots v_k(\tilde t)|0\rangle=2\langle 0|v_1(\tilde t)v_2(\tilde t)\cdots v_k(\tilde t)\phi_0 |0\rangle
\]
and again we find a Pfaffian.
We thus arrive at the main theorem.
\begin{theorem} \label{Theo reduced} 
All polynomial tau-functions of the $n$-th Sawada-Kotera hierarchy are, up to a scalar
factor,     equal to the Pfaffian
\begin{equation}
\label{TBK reduced}
\tau_\lambda (\tilde t)=Pf\left(\chi_{\lambda_i, \lambda_j}(\tilde t+c_{\overline\lambda_i},\tilde t+c_{\overline\lambda_j})\right)_{1\le i,j\le 2m},
\end{equation}
where $\lambda=(\lambda_1,\lambda_2,\ldots ,\lambda_{2m})$, $m=0,1,\ldots$,   is an extended strict partition,  which satisfies the $-n$ shift condition   \eqref{condition V lambda} for \eqref{V lambda}.  The polynomials  $ \chi_{a,b}$ are given by \eqref{chi}.
Here, as before,  $\tilde t=(t_1,0,t_3, 0,\ldots)$,  and 
$c_{\overline \lambda_i}=(c_{1,\overline \lambda_i},c_{2,\overline \lambda_i}, c_{3,\overline \lambda_i}, \ldots)$ are arbitrary constants, where we replace, recursively,  for all $j=1,2,\ldots 2m$ (respectively for all $j=1,2,\ldots 2m-1$), when $\lambda_{2m}\ne 0$ (resp.  $\lambda_{2m}=0$), the constants $c_{\lambda_j+\lambda_\ell, \overline\lambda_j}$,  for $j\le \ell\le 2m$  (resp.   $j\le \ell<2m$)  as follows:\\
(1) if $\overline\lambda_j\ne \overline\lambda_\ell$, then 
\begin{equation}
\label{restrict1}
\begin{aligned}
&c_{\lambda_j+\lambda_\ell, \overline\lambda_j}=
-(-1)^{\lambda_j+\lambda_\ell}\times\\
&\quad s_{\lambda_j+\lambda_\ell}(
c_{1,\overline\lambda_\ell}-c_{1,\overline\lambda_1},c_{2,\overline\lambda_\ell}+c_{2,\overline\lambda_1},
\ldots,
	c_{\lambda_j+\lambda_\ell-1,\overline\lambda_\ell}+(-1)^{\lambda_j+\lambda_\ell-1}c_{\lambda_j+\lambda_\ell-1,\overline\lambda_j},c_{\lambda_j+\lambda_\ell, \overline\lambda_\ell})	.
	\end{aligned} .
\end{equation}
(2) If $\overline\lambda_j= \overline\lambda_\ell$ and $\lambda_j+\lambda_\ell$ is even, then 
\begin{equation}
\label{restrict2}
c_{\lambda_j+\lambda_\ell, \overline\lambda_j}=-\frac12s_{\frac{\lambda_j+\lambda_\ell}2}(2c_{2,\overline\lambda_j},2c_{4,\overline\lambda_j},
\cdots, 2c_{\lambda_j+\lambda_\ell-2},0).
\end{equation}
\end{theorem}
{\bf Proof.}
We still need to use the fact that all vectors $v_i$, for $i=1,\ldots ,k$ form an isotropic subspace. 
So assume that $1\le j\le \ell\le k$  and that $v_j$ and $v_\ell$ are given by \eqref{v lambda4}.  Then 
\begin{equation}
\label{restrict isotropic}
\begin{aligned}
0=(v_j,v_\ell)=& (-1)^{\lambda_j}\sum_{i=0}^{\lambda_j+\lambda_\ell} (-1)^is_i(c_{\overline\lambda_j})s_{\lambda_j+\lambda_\ell-i}(c_{\overline\lambda_\ell})\\
=&(-1)^{\lambda_j}s_{\lambda_j+\lambda_\ell}(c_{i,\overline\lambda_\ell}+(-1)^ic_{i,\overline\lambda_j}).
\end{aligned}
\end{equation}
Here we have used the fact that the coefficient of $z^m$ of 
\[
e^{\sum_{i=1}^\infty x_iz^i+y^i(-z)^i} = e^{\sum_{i=1}^\infty x_iz^i} e^{\sum_{i=1}^\infty  y^i(-z)^i}
\]
is equal to  $s_m(x_i+(-1)^i y_i)=\sum_{j=0}^m (-1)^{m-j}s_j(x)s_{m-j}(y)$.

So we need to investigate  condition  \eqref{restrict isotropic}.  Here we have two possibilities, viz.  $\overline\lambda_j\ne \overline\lambda_\ell$ and 
$\overline\lambda_j= \overline\lambda_\ell$.  If $\overline\lambda_j\ne \overline\lambda_\ell$, then using the fact that $s_i(x)=x_i+$ terms not containing $x_i$, we find \eqref{restrict1}.
If $\overline\lambda_j= \overline\lambda_\ell$, then  notice that $s_{\lambda_j+\lambda_\ell}$ only depends on the $c_{i,\overline\lambda_j}$ with $i$ even. Thus all elementary Schur polynomials  $s_{2i+1}$,  in only the even variables are  equal to zero.  This means that if $\lambda_j+\lambda_\ell$ is odd, there is no restriction on the constants,  but if $\lambda_j+\lambda_\ell$ is even, we find that
\[
c_{\lambda_j+\lambda_\ell, \overline\lambda_j}=
-\frac12s_{\lambda_j+\lambda_\ell}(0,2c_{2,\overline\lambda_j},0,2c_{4,\overline\lambda_j},0,2c_{6,\overline\lambda_j},\ldots, 0,2c_{\lambda_j+\lambda_\ell-2, \overline\lambda_j},0,0)
 .
\]
Note that this   restriction coincides with \eqref{restrict2}.
\hfill$\square$
\begin{remark}
Since 
$\chi_{\lambda_j,\lambda_\ell}$ is given by \eqref{chi},  the constant
 $c_{2\lambda_1,\overline\lambda_1}$ does not appear in \eqref{TBK reduced} and the  substitution \eqref{restrict2} for  
 $c_{2\lambda_1,\overline\lambda_1}$
 is void.
\end{remark}
For $n=3$,  see Example \ref{Ex}(b),  we only have one infinite series of constants $c_{i,\overline\lambda_1}$, which means that we only have the substitutions \eqref{restrict2}.    We therefore find:
\begin{corollary}
\label{cor}
All polynomial tau-functions of the
Sawada-Kotera hierarchy are, up to a non-zero constant factor, 
\begin{equation}
\tau_\lambda (\tilde t)=Pf\left(\chi_{\lambda_i,\lambda_j}(\tilde t+c ,\tilde t+c )\right)_{1\le i,j\le 2m},
\end{equation}
where   $\lambda$ is   one of the following extended strict partitions:
\begin{enumerate}
\item 
$(6m+1, 6m-2, 6m-5,\ldots, 4,1)$;
\item 
$(6m-2, 6m-5, 6m-8,\ldots, 4,1,0)$;
\item
$(6m+2,6 m-1, 6m-4,\ldots, 5,2)$;
\item
$(  6m-1, 6m-4,6m-7,\ldots, 5,2,0)$,
\end{enumerate}
where
$m=0,1,\ldots$, 
and  $c=(c_1,c_2,c_3,\ldots) $   are arbitrary constants in which we substitute recursively, $c_2=0,$  and $c_8,  c_{14}, c_{20},\ldots, c_{12m-4}$ in case 1;
 $c_2=0$,  and $c_8,  c_{14}, c_{20},\ldots, c_{12m-10}$ in case 2; $c_4,  c_{10}, c_{16},\ldots, c_{12m-2}$ in case 3; and $c_4,  c_{10}, c_{16},\ldots, c_{12m-8}$ in case 4,  respectively,  by the following formula
 \begin{equation}
\label{restrict3}
c_{2k}=-\frac12s_{k}(2c_2 ,2c_4, \ldots 2c_{2k-2},0) \quad \mbox{for }k>1.
\end{equation}
\end{corollary}  
\begin{remark}
If we choose  $c=\tilde c=(c_1,0, c_3,0,c_5,\ldots)$,  then  the corresponding Sawada-Kotera tau-function $\tau_\lambda(\tilde t)$   is equal,  up to a
non-zero constant 
 factor, to a Schur $Q$-function
$Q_\lambda(\tilde t+\tilde c)$ (cf. \cite{You}).
\end{remark}
\begin{example} All the Sawada-Kotera tau-functions related to the partition $\lambda=(5,2)$ are given, up to a multiplicative constant,   by
\[
\begin{aligned}
&c_1^7 + 1120 c_7 + 7 c_1^6 t_1 - 280 c_2^3 t_1 - 
 280 c_5 t_1^2 + 140 c_2^2 t_1^3 + t_1^7 + 
+ 35 c_1^3 (2 c_2 + t_1^2)^2 +
\\
&+ 7 c_1t_1^5 (2 c_2 + 3 t_1^2) + 
 35 c_1^4 (2 c_2 t_1 + t_1^3) - 280 t_1^2 t_5 - 
 7 c_1^2 (40 c_5 - 60 c_2^2 t_1 - 20 c_2 t_1^3 - 3 t_1^5 + 
    40 t_5) + \\
    &+14 c_2(120 c_5 + t_1^5 + 120 t_5) - 
 7 c_1 (40 c_2^3 - 60 c_2^2 t_1^2 - 10 c_2 t_1^4 + 
    t_1(80 c_5 - t_1^5 + 80 t_5)) + 1120 t_7,
    \end{aligned}
\]
with    $c_i\in \mathbb C$, arbitrary.
(We have eliminated $c_4$ by the substitution $c_4=-c_2^2$,  all the other constants that do not appear, disappear automatically.)
\end{example}

\end{document}